\documentclass[twocolumn,showpacs,preprintnumbers,amsmath,amssymb]{revtex4}
\usepackage{graphicx}
\usepackage{dcolumn}
\usepackage{bm}
\usepackage{amssymb}
\usepackage{epsfig}
\newcommand{\be}{\begin{equation}}
\newcommand{\ee}{\end{equation}}
\newcommand\beq{\begin{eqnarray}}
\newcommand\eeq{\end{eqnarray}} 
\newcommand\eqn[1]{\label{eq:#1}} 
\newcommand\eq[1]{eq. (\ref{eq:#1})} 
\newcommand{\vev}[1]{\langle #1 \rangle}
\newcommand{\bfk}{{\mathbf k}}

\newcommand{\GeV}{{\rm ~GeV }}

\newcommand{\CN}{{\cal N}}
\newcommand{\CO}{{\cal O}}
\newcommand{\CQ}{{\cal Q}}

\begin{document}

\preprint{INT-PUB-08-32}
\preprint{IFT-UAM/CSIC-08-51}

\title{Inflationary Axion Cosmology Beyond Our Horizon}
\author{David B. Kaplan$^{1}$}
\email{dbkaplan@phys.washington.edu}

\author{Ann E. Nelson$^{2}$}
\email{anelson@phys.washington.edu}
\affiliation{$^1$Institute for Nuclear Theory, University of Washington, Seattle, WA 98195-1550,USA}
\affiliation{$^2$Dept. of Physics, University of Washington,  Seattle, WA 98195-1560,USA}

\begin{abstract}
 In theories of axion dark matter with large axion decay constant, temperature variations in the CMB are extremely sensitive to perturbations in the initial axion field,  allowing one to place a lower bound on the total amount of inflation.  The most stringent bound comes from axion strings,  which  for axion decay constant $f_a=10^{17}\GeV$  would currently be observable at a distance of $6\times 10^{16}$ light-years, nearly $10^7$ times as far away as our horizon.
 \end{abstract}
\date{\today}
\maketitle

\section{Introduction}

The  Peccei-Quinn (PQ) solution \cite{Peccei:1977hh,Peccei:1977ur} to the strong CP problem posits a $U(1)$ symmetry which is exact up to a QCD anomaly and nonlinearly realized below a temperature $T_{PQ}$.  This implies the existence of a light pseudo Goldstone boson --- the axion \cite{Weinberg:1977ma,Wilczek:1977pj} --- whose low energy properties are largely determined in terms of its decay constant $f_a\simeq T_{PQ}$ and an integer $\CN$ characterizing the strength of the PQ symmetry's QCD anomaly \cite{Kim:1979if,Shifman:1979if,Zhitnitsky:1980tq,Dine:1981rt}.   The decay constant $f_a$ is defined so that $\theta(x)=a(x)/f_a$ is an angle taking values  $\theta\in [0,2\pi)$, where $a(x)$ is the canonically normalized axion field. To leading order in chiral perturbation theory, the axion mass is 
$
m_a
 \approx 6\times 10^{-9}\text{ eV}  \times f_{16}^{-1} 
$, 
 where 
  $f_{16} \equiv (f_a/\CN)/(10^{16}\GeV)$. Laboratory and astrophysical considerations give $f_{16}\gtrsim 10^{-6}$, a bound involving model-dependent axion couplings to photons and electrons \cite{Kaplan:1985dv,Srednicki:1985xd}.
 
  Cold relic axions from the big bang are a viable candidate for the cosmic dark matter \cite{Preskill:1982cy,Abbott:1982af,Dine:1982ah,Sikivie:2006ni}.  The PQ symmetry ensures that for temperature $\Lambda_{QCD}\ll T \lesssim T_{PQ}$ the free energy of the universe is independent of the angle $\theta$; as a result, the universe will be populated by topological defects in the form of axion cosmic strings, around which $\theta$ varies from zero to $2\pi$.  Far from these strings, $\theta(x)$ will assume some random initial value $\theta_i$, with spatial fluctuations on scales less than the horizon damping out with the cosmic expansion. When the temperature falls to $T\sim \Lambda_{QCD}$, the QCD anomaly lifts the vacuum degeneracy in $\theta$ and there arise $\CN$ local minima, whose zero temperature curvature is given by the axion mass.     In general, the initial value $\theta_i$ will not be aligned with a minimum of the potential (taken to lie at $\theta=0$), and coherent axion oscillations will result. For small $\theta_i$ the relic energy density today stored in these coherent oscillations is \cite{Fox:2004kb,Beltran:2006sq}
\beq
\Omega_a \simeq \Omega_c (\CN \theta_i)^2 \times
 \begin{cases}
5\times 10^{5} f_{16}^{7/6}\,, & f_{16}<1/10 \cr
3\times 10^{4} f_{16}^{3/2}\,, & f_{16} > 10
\end{cases}
\eqn{fbd}
\eeq
where  $\Omega_{\text{c}}\simeq 0.23 $  is the cold dark matter that fits observation in the  $\Lambda$CDM model \cite{Komatsu:2008hk}.
The above expression ignores nonlinearities in $\theta_i$, but one can safely assume that for large $\theta_i$, the factor of $\theta_i^2$ in the above expressions is replaced by a number of O(1).
The uncertainty in \eq{fbd} and the absence of a formula for $1/10 \lesssim f_{16} \lesssim 10$ are due to the crude understanding of the temperature dependence of the axion mass which arises from nonperturbative QCD effects when the temperature is $T\simeq \Lambda_{QCD}$.  In principle, this ignorance could be remedied in part by lattice calculations.   Nevertheless, it is apparent that (i) if $ \theta_i\sim O(1)$ then $f_a\simeq 10^{12}\GeV$ would allow axions to be the dark matter today; (ii) if $f_a\gg 10^{12}\GeV$ then $\theta_i\ll 1$ is required; (iii) for $f_{16}\simeq 1$, interpolation between the two formulas in \eq{fbd} suggests $\CN \theta_i \simeq 2\times 10^{-3}$ would give the correct dark matter density.

If the universe does not undergo inflation after the universe is in thermal equilibrium at a temperature below $T_{PQ}\sim f_a$, then causality arguments imply that $\theta_i^2\sim O(1)$, which means that one must have $f_a\lesssim 10^{12}\GeV$ or else axions would provide a larger value for $\Omega_{\text{dm}}$ than observed.  The ADMX experiment is currently in the process of probing this interesting ``axion window",  $10^{10}\GeV\lesssim f_{a} \lesssim 10^{12}\GeV$ \cite{Asztalos:2003px}.

Neutrino masses, the possible unification of the couplings of the standard model, and the so-called ``model-independent" axion in string theory \cite{Gaillard:2005gj,Svrcek:2006yi} all suggest one should consider a higher decay constant, such as $f_a\sim 10^{16}\text{ GeV}$.  By \eq{fbd} this would imply an initial $\theta_i$ which is very small over a region of the universe that covers our horizon today, which is possible if inflation occurs and the reheat temperature is below $T_{PQ}$ \cite{Pi:1984wk}.  Having  such a finely tuned initial condition for $\theta_i$ might be explained by  the anthropic principle  \cite{Linde:1987bx,Turner:1990uz,Linde:1991km,Tegmark:2005dy}, the basic argument being that all possible values of $\theta_i\in [0,2\pi)$ occur with equal probability in the universe, but only in those patches with small $\theta_i$ are habitable galaxies produced. 

We will call this the ``ultralight axion scenario": an axion with $f_a \gg 10^{12}\GeV$, a period of inflation below the $PQ$ scale to ensure a homogeneous value of the initial axion misalignment angle $\theta_i$ over today's observable universe, and an anthropic explanation for its apparent fine-tuning.

It is  important to obtain  evidence for or against  the ultralight axion. In the absence of methods for direct detection of such weakly interacting particles, it is natural to seek cosmological signatures.  Precisely because $\theta_i$ is so finely tuned, the dark matter density today is very sensitive to small spatial fluctuations in the axion field, $\delta\theta_i(x)$. In this Letter we discus three different types of ultralight axion fluctuations which could survive into a post-inflationary Universe and could be detectable today.

\section{Axion fluctuations}

Axion fluctuations are isocurvature, arising before the energy density of the universe depends on $\theta$.  The effect of isocurvature fluctuations on the CMB spectrum is greatest at low  $\ell$, producing  acoustic peaks in the CMB of much less power than the quadrupole and out of phase with what is observed (see \cite{Fox:2004kb}).  Observation agrees well with an initial scale invariant adiabatic spectrum; normalizing such a spectrum to the first acoustic peak in the CMB, the predicted quadrupole moment is in agreement with the COBE result \cite{Bennett:1996ce}, and hence places the constraint that a flat spectrum of  isocurvature perturbations comprise $\le 0.067$ of the total perturbation spectrum \cite{Komatsu:2008hk}.    This constraint has been applied to inflationary axion perturbations, and results in an upper bound on the inflationary Hubble constant $H_I$ for large $f_a$   \cite{Seckel:1985tj,Linde:1985yf,Burns:1997ue,Kawasaki:1997ct,Turner:1990uz,Lyth:1989pb,Lyth:1991ub,Lyth:1992tx,Fox:2004kb,Kain:2005ic,Beltran:2006sq,Hertzberg:2008wr}, as we recapitulate below.

 The total fluctuation spectrum was fixed by COBE \cite{Bennett:1996ce}  to satisfy $(\langle(\delta T^2\rangle)^{\frac12}\approx 1.1 \times 10^{-5} T \ ,$ assuming a scale invariant spectrum. COBE was primarily sensitive to the low $\ell$ modes  in the fluctuation spectrum.

When the quadrupole constraint is satisfied, the $\ell>2$ contributions from axion fluctuations to the CMB are small and provide no new constraints.  However, it is interesting to consider the monopole and dipole contributions.  Dipole isocurvature effects can lead to observable consequences -- the tilted universe scenario of Turner \cite{Turner:1991dn} where the rest frame of the CMB is not the same as the matter rest frame.  As a consequence, measurements of the peculiar velocity of the earth relative to the CMB will not coincide with the velocity relative to distant supernovae.  The constraints on the isocurvature dipole moment are currently an order of magnitude weaker than the constraints from the quadrupole moment.  Thus  the dipole could only be observable  when it  is enhanced relative to the quadrupole, such as by the geometric enhancement $(k/k_\text{hor})^\ell$ which occurs when superhorizon  modes  with $k< k_\text{hor}$ dominate \cite{superhorizon}.  This can arise  from both thermal or topological fluctuations.

Monopole fluctuations have been cited as making a positive definite contribution to $\Omega_a$, and hence subject to constraints \cite{Turner:1990uz,Fox:2004kb}. The argument is that $\theta_i^2$ is replaced by $ (\theta_i^2 + \sigma^2)$ in the relic axion energy relation \eq{fbd}, where $\sigma^2$ is the fluctuation contribution. This replacement would signify that inflation-induced fluctuations make a positive-definite contribution to the relic axion energy density today, regardless of how finely-tuned $\theta_i$ was prior to inflation.  However this is incorrect --- no matter how much  the average value of $\vev{\theta}$  within a particular volume may wander  during inflation, after inflation all values of theta are still equally likely, and there are always regions of the universe where $\vev{\theta}$ is acceptably small --- and the anthropic argument then implies that our universe today must be located in  such a region. Therefore the monopole fluctuation in the axion field has no observable consequence.

\subsection{Scale invariant fluctuations}

The first type of ultralight axion fluctuation we consider are those induced during inflation, and we reproduce a bound in the literature.   
 During inflation the axion field develops a perturbation spectrum for modes crossing the causal horizon of the form
\beq
\vev{ \delta a_{\bfk} \delta a^*_{\bfk'} } = \frac{2\pi^2}{k^3} \left(\frac{H_k}{2\pi}\right)^2 (2\pi)^3 \delta^3(\bfk - \bfk')\ ,
\eqn{cor}
\eeq
where $H_k$ is the value of the Hubble parameter when the mode of wavenumber $k$ crosses the horizon.  For slow roll inflation with a nearly $k$-independent Hubble parameter $H_k\simeq H_I$, fluctuations in the axion field \eq{cor} lead to  $\ell\ge 1$ multipole moments in $\theta_i$  dominated by modes with wavelength of order the horizon, with amplitude
$
|\delta\theta|_\ell\approx H_I/2\pi f_a
$.

 Following the calculations of ref. \cite{Fox:2004kb}, the axion contribution to the total fluctuation spectrum is
 \beq\left.\frac{\delta T}{T}\right|_{\rm axion}=-\frac{12}{15}\frac{\Omega_a}{\Omega_c}\frac{\delta\theta}{\theta_i}\ , \eeq
 which, when combined with the COBE and WMAP  bounds implies 
 \beq\frac{\Omega_a}{\Omega_c}\frac{\delta\theta}{\theta_i}< \frac{15}{12} (1.1 \times 10^{-5})(0.067) =9.2\times 10^{-7}\ .\eqn{isocurvbound}\eeq

Assuming that ultralight axions constitute all of  the dark matter, and are not diluted by late epoch entropy production, the constraint on $\delta\theta$ implies  a bound on $H_I$
\beq
H_I \lesssim H_I^{max}\equiv\begin{cases}
f_{16}^{5/12}\times 5\times  10^8\GeV, & f_{16}\lesssim 10^{-1} \cr
f_{16}^{1/4} \ \times3\times 10^{8}\GeV, & f_{16}\gtrsim 10  \end{cases}
\eqn{hbound}
\eeq
(For intermediate values of $f_a$ we interpolate between these bounds.)  Inflation at such low scales  is inconsistent with the simplest single field inflation models \cite{Pagano:2007st}. 
 Observation of tensor modes in the CMB from an inflationary spectrum of gravitational waves will be possible in future experiments \cite{Bock:2006yf}  provided $H_I$ is high,  about  $10^{13}$ GeV, which would rule out the ultralight axion scenario unless $f_a\gg M_p$;  such a high value for $f_a$  has been argued to  be inconsistent with  string theory \cite{Banks:2003sx,Ooguri:2006in,Svrcek:2006yi}.

   \subsection{Scale dependent fluctuations }
  
Evidence for scale dependent axion fluctuations would be interesting, giving us information about inhomogeneities preceding inflation.  Inflation was discovered as a means  to make  information about initial conditions   inaccessible to the contemporary observer, but ultralight axions---being fine-tuned by the anthropic principle---provide a probe that is more sensitive to initial conditions than the observed flatness of the universe or the smallness of $\delta T/T$ in the CMB spectrum in generic $\Lambda$CDM models.   The question is:  how completely did inflation erase information about prior inhomogeneities?

Two length scales upon which the axion fluctuation spectrum could depend are the size of the patch destined to become our horizon today, and the thermal wavelength at the beginning of inflation.  (Other scales can arise from non-thermal primordial axion fluctuations --- we will consider those of topological origin below ---or from nontrivial time dependence of the Hubble scale during inflation, which we do not discuss).
We take $N_\text{min}$ to be the minimum number of $e$-foldings required so that our entire horizon today was causally connected at the beginning of inflation,  and $N$ to be the total number of $e$-foldings that actually occur during inflation, with $N\ge N_\text{min}$.  Then the quantity
 \be R_X\equiv e^{N-N_\text{min}} \ee
has a geometric interpretation as the ratio of the horizon size at the start of inflation, to the radius of the patch destined to become our present day horizon.  Fluctuations of the size of our horizon have comoving wavenumber 
\beq
k_\text{hor}=R_X H_I, 
\eqn{khor}\eeq
where $H_I$ is the Hubble parameter at the beginning of inflation.   Equivalently, $R_X$ times our current horizon size gives the present size of the local patch of the universe which was  in causal contact at the start of inflation.  

Assuming that the preinflationary universe was not especially flat or uniform,  observations can provide lower limits on $R_X$.   For inflation to explain the observed flatness of the universe,  assuming that the curvature radius was $\CO(1)$ times the Hubble radius at the beginning of inflation, requires $R_X\simeq \vert\Omega-1\vert^{-1/2} \gtrsim 10$, given limits on $\Omega$  from WMAP data \cite{Komatsu:2008hk} . To avoid an unacceptably large  CMB quadrupole moment arising from generic $\CO(1)$ preinflationary classical field fluctuations  requires $R_X \gtrsim 200$  \cite{Turner:1991dn}.  Here we are interested in how ultralight axion models lead to constraints on $R_X$, or the existence of an ultralight axion might be inferred from observation.

\subsubsection{Preinflationary thermal fluctuations}

We next consider observable effects due to preinflationary thermal fluctuations in the axion field, assuming that thermal equilibrium existed  at a temperature $T\sim T_{PQ}$.  As Goldstone bosons, axion interactions fall out of thermal equilibrium rapidly below that temperature, after which their spectrum follows a redshifted thermal distribution for each mode until it crosses the horizon during inflation.  To discuss preinflationary thermal fluctuations, it is convenient to define the quantity $R_T$ in terms of the temperature $T_I$ at the start of inflation, where
\beq
T_I=R_T H_I . 
\eqn{rt}\eeq
Parametrically we expect $H_I \sim T_I^2/M_P$.

In terms of the comoving wavenumber $k$ and cosmological scale factor $R(t)$, the correlation between axion modes  with $H_I < k < T_I$  is given by 
\be
 \langle \delta a^*_{\bf  {k}} \delta a_{\bf  {k}'}\rangle\simeq \frac{T_I}{R(t)^2k^2} (2\pi)^3\delta^3({\bf  k}-{\bf k'}), 
 \eqn{therm}
 \ee
We have defined the scale factor  $R(t)$ to equal one at the start of inflation.  For $k>T_I$  Boltzmann suppression exponentially damps the thermal fluctuations, leaving only the zero-temperature scale invariant fluctuations discussed in the previous section.

 After inflation has begun, axion fluctuations cross the horizon at time $t_k$ where $k/R(t_k) = H(t_k)$. In models of slow-roll inflation $H(t_k)\simeq H_I$ and so the thermal fluctuations that we eventually see  are given by 
 \be
 \langle \delta a_{\bf  {k}} \delta a_{\bf  {k}'}\rangle\simeq \frac{T_I H_I^2}{k^4} (2\pi)^3\delta^3({\bf  k}-{\bf k'}), 
\qquad\text{(slow roll).} \eqn{froz}
 \ee
Note that these fluctuations are parametrically enhanced by a factor of $T_I/k\gtrsim 1$ relative to the inflation-induced fluctuations in  \eq{cor}.  Again, they are isocurvature fluctuations and are most important in the  dipole and quadrupole moments.

It is convenient to directly compare the contributions of the thermal fluctuations \eq{therm} to low multipole moments of the RMS energy density $|\delta\rho/\rho| = 2|\delta\theta/\theta|$ to the contributions from the zero temperature fluctuations in \eq{cor}.
For $\ell=0,1,2$ the ratio of thermal to inflationary contributions is

\beq
\frac{(\delta \rho/\rho)_{\ell,\text{therm.}}}{(\delta \rho/\rho)_{\ell,\text{inf.}}} \sim \begin{cases} \sqrt{R_T/R_X} & R_X\le R_T\cr  (R_T/R_X)^\ell & R_X\ge R_T \end{cases}\ .
\eqn{pratio}
\eeq

For $R_T < R_X$ the thermal fluctuations have a UV cutoff corresponding to a length scale $T_I^{-1}$, which is larger than the size of our horizon patch,  $k_\text{hor}^{-1}$, explaining why their effect falls off as $(R_T/R_X)^\ell= (T_I/k_\text{hor})^\ell$.  For $R_T>R_X$ the size of our horizon patch is effectively the UV cutoff for fluctuations contributing to large scale anisotropies.  Therefore the sole temperature dependence comes from the factor of $T_I$ in amplitude of the fluctuations \eq{therm}, which enters as $\sqrt{T_I} $ in the RMS energy fluctuation, explaining the $\sqrt{R_T/R_X}$ behavior.

Thus there is a bound from the quadrupole moment on thermal fluctuations analogous to the bound \eq{hbound} on inflationary fluctuations, found by substituting $H_I\to H_I\sqrt{R_T/R_X}$. By the relation $H_I\sim T_I^2/M_p$, the bound can be expressed as the inequality
\beq
R_X \gtrsim \sqrt{\frac{M_p}{H_I^{max}}} \, \left(\frac{H_I}{H_I^{max}}\right)^{3/2}\ .
\eqn{tbound}
\eeq 
If, for example, we take $f_{16} = 1/10$ and $H_I = H_I^{\max}= 2\times 10^8\GeV$, 
then we get the nontrivial bound from the quadrupole moment $R_X \gtrsim  3\times 10^5$.  If instead $H_I =10^6\GeV$, then the bound is weaker, $R_X\gtrsim 100$.

\subsubsection{Topological disorder}

Cosmic axion strings formed at the PQ phase transition provide a source of nonthermal scale-dependent axion fluctuations.  As the universe subsequently expands  these strings  intersect and detach loops which decay into axions. Causality dictates that there should be $\CQ \gtrsim O(1) $  strings traversing each horizon volume at later epochs.   These give rise to gradients in the classical axion field, with the variation $\Delta\theta$ across our horizon patch being 
\beq
\Delta\theta \gtrsim 2\pi\CQ\frac{H_I}{k_\text{hor}} = \frac{2\pi\CQ}{R_X}\ .
\eeq
giving rise to a dipole moment in the energy density of order $\gtrsim 2\pi\CQ/(\theta_i R_X)$.  Higher multipole moments are suppressed by $R_X^{-\ell}$.  The inequality arises because we could be accidentally close to an axion string. During inflation and the subsequent evolution of the universe, $\Delta\theta$ and the dipole contribution to the axion fluctuation spectrum is unchanged.
The presence of $\theta_i$ in the denominator of this expression gives a  strong constraint on $R_X$ for large axion decay constant.
From \eq{fbd}   
we arrive at the estimate for  the dipole asymmetry $q_1$
\beq q_1\sim 
\frac{2\pi\CQ}{R_X}\left(\frac{\Omega_a}{\Omega_c}\right)^{\frac12}  \times\begin{cases}  
6.5\times 10^2 f_{16}^{{7/12}}, & f_{16} < 10^{-1} \cr
1.8 \times 10^2\ f_{16}^{3/4} , &    f_{16} > 10 .  
\end{cases}
\eeq

A dipole asymmetry of order $10^{-3}$ is observed in the CMB temperature \cite{Kogut:1993ag}, and is usually assumed to be due to the peculiar velocity of the solar system with respect to the CMB. 
An isocurvature dipole moment, however, causes matter to flow toward the region of higher dark matter density so that the rest frames of matter and the CMB do not coincide \cite{Turner:1991dn}.   Therefore one might expect our velocity relative to distant supernovae to differ from the peculiar velocity deduced from the CMB dipole, which is of order $10^{-3}$. Various techniques are used to determine the average velocity of galaxies relative to the CMB at cosmological scales
 \cite{Langlois:1995ca,Strauss:1995fz,Kocevski:2004pf,Cooray:2006ft,Haugboelle:2006uc,Gordon:2007sk},   It has been estimated that such an effect could be detected down to of order 10\% 
of the observed dipole with future analysis of large scale cosmological surveys \cite{Gordon:2007sk}. For axion dark matter with $f_{16}\sim 10$, a constraint on the isocurvature dipole $q_1$ at the $10^{-4}$ level would be sensitive to the presence of a cosmic string  more than $ 10^7$ horizon lengths away.

 \section{Conclusions}
An ultralight axion is a well motivated dark matter candidate, provided  inflation takes place at a sufficiently low scale.  In this paper we have discussed how  preinflationary thermal and topological fluctuations of the axion give rise to isocurvature quadrupole and dipole moments of the primordial perturbation spectrum, providing constraints on---and potentially providing indirect evidence for---dark matter, inflation, the preinflationary universe, a very high physical scale,  anthropic selection of the dark matter abundance, and the explanation for the small electric dipole moment of the neutron!  Note that similar calculations may be done for any light, weakly coupled scalar field, such as a string modulus.

\section*{Acknowledgments}

We would like to thank the Instituto de Fisica T\'eorica of the Universidad Aut\'onoma de Madrid for hospitality.
This work was supported in part by the DOE under contracts DE-FG03096-ER4095 and DE-FGO3-00ER41132, 
by the Spanish MEC
grant SAB2006-0089 and project FPA2006-05423, and the regional «Comunidad
de Madrid« HEPHACOS project.

\bibliography{axicosmo}
\bibliographystyle{apsrev}

\end{document}